# Proper Characterization of Heat-to-Electric Conversion Efficiency of Liquid Thermogalvanic Cells

Qiangqiang Huang[1,2], Yuchi Chen[1], Ronggui Yang[1, 3, *], and Xin Qian[1, 2*]

[1] School of Energy and Power Engineering, Huazhong University of Science and Technology, Wuhan 430074, China.

[2] State Key Laboratory of Coal Combustion, Huazhong University of Science and Technology, Wuhan 430074, China.

[3] School of Engineering, Peking University, Beijing 100871, China.

*Corresponding emails: xinqian21@hust.edu.cn; ronggui@pku.edu.cn

## ABSTRACT

Liquid thermogalvanic cells (LTCs) have recently emerged as a promising approach for harvesting low-grade heat, owing to their low cost, compact structure, and intrinsic high thermopower. However, significant discrepancies exist in quantifying the output power and the heat-to-electric energy conversion efficiency. While the figure of merit $Z_{el}T = S_{tg}^2 \sigma T/\kappa$ is often reported by separately characterizing the thermopower $S_{tg}$, ionic conductivity $\sigma$, and thermal conductivity $\kappa$ of the electrolyte, this material-property-based $Z_{el}T$ failed to include the electrochemical reaction kinetics at the electrolyte-electrode interface, which is the key process determining the harvested power and the irreversible losses. This work establishes an experimental protocol for accurately characterizing the efficiency of LTCs. We proposed a proper figure of merit $ZT = S_{tg}^2 T/R_{in}K$ for LTC devices, where $R_{in}$ and $K$ are the total internal resistance and thermal conductance across the LTC. This $ZT$ of the device is derived by linearizing the Butler-Volmer relation, which incorporates the irreversible losses such as mass transfer and activation overpotential near the electrode surfaces. Different methods for characterizing the output power of LTCs are also discussed thoroughly, including linear sweeping voltammetry (LSV), constant resistance discharging, and constant current steps discharging. LSV tends to overestimate the output power and efficiency due to the transient effects, while constant current steps and constant resistance discharging give reasonable output power at the steady state. Heat conduction across the LTC is also carefully measured, and the natural convection of electrolytes makes a considerable parasitic contribution to heat conduction across the device. With careful experimental characterization, we showed that the modified $ZT = S_{tg}^2 T/R_{in}K$ is a proper efficiency indicator, provided that the steady state is ensured during measurements.

I. INTRODUCTION

Liquid thermogalvanic cells (LTCs) offer a cost-effective and compact method for converting low-grade heat into electricity. The effective thermopower $S_{tg}$ of LTCs can be defined as $S_{tg} = -\Delta E/\Delta T$, with $\Delta E$ and $\Delta T$ the electrochemical potential difference and the temperature difference across the device. In liquid electrolytes, $S_{tg}$ is generally determined by the reaction entropy change $S_{tg} \propto \Delta s_{rxn}$ of redox ion pairs [1,2], which is much larger than the entropy of transport along with electrons and holes in semiconductors [3], giving rise to much larger thermopower (~1 mV/K for aqueous electrolytes) of LTCs than semiconductors (~10 µV/K) [4-6]. Recent five years have witnessed significant advancements in further boosting $S_{tg}$ by engineering solvation structure or activity of redox ion pairs. Two- or threefold increases in p-type ($S_{tg} > 0$) and n-type ($S_{tg} < 0$) thermopower have been achieved, including the -3.6 mV/K for $Fe^{3+}/Fe^{2+}$ in acetone [7,8] and 3.7 mV/K for $Fe(CN)_6^{4-}/Fe(CN)_6^{3-}$ electrolytes with selective crystallization of $Fe(CN)_6^{4-}$ with guanidinium ions [9]. These advancements have enabled impressive Carnot-relative efficiency up to ~11%, harvesting low-grade heat sources below 100 °C.

Despite these progresses, there is still no consensus on how to quantify the energy conversion efficiencies of LTCs. Since LTCs operate similarly to a solid-state thermoelectric generator (TEG), a figure of merit $Z_{el}T = S_{tg}^2 \sigma T/\kappa$ can be defined analogously and is often reported as an efficiency indicator of LTCs [9-13], where $\sigma, \kappa$ are the ionic conductivity and thermal conductivity of the electrolyte. However, there are significant differences in the heat-to-electric conversion mechanisms between LTCs and TEG, which makes directly using $Z_{el}T$ for evaluating efficiency of LTCs questionable. In TEGs, the temperature gradient $\nabla T$ drives electrons or holes migrating across the material and induce an internal electric field $\vec{\mathcal{E}}$. Therefore, the conventional Seebeck coefficient $S$ can be well-defined at each point $x$ within the material as $\vec{\mathcal{E}}(x) = S\nabla T(x)$ [3]. In LTCs, however, the voltage is dominantly generated by the electrochemical potential difference $\Delta E$ when the two electrodes are held at different temperatures. $\Delta E$ is determined by the local electrochemical equilibrium and local temperature at the electrode-electrolyte interfaces, rather than $\nabla T$ within the bulk electrolyte [9]. Therefore, the effective thermopower $S_{tg}$ of an LTC cannot be understood as a nonequilibrium Seebeck coefficient [14]. During discharging, the electrochemical reactions take place at the electrode-electrolyte interfaces, such that the mass transport and reaction kinetics are key processes affecting the output power and efficiency [15]. These interfacial effects are not included when the figure of merit $Z_{el}T$ is naively defined using electrolyte properties.

Another way of evaluating the efficiency is by taking the ratio between the harvested power and the absorbed heat from the reservoir. Unfortunately, inconsistent methods have been used to determine the output discharging curves and the harvested power. For example, linear sweep voltammetry (LSV) is often used to determine the maximum power of an LTC, where the voltage is linearly decreased from the open circuit voltage to zero and the responsive current is simultaneously measured. However, inconsistent scanning rates from a few mV/s to dozens of mV/s have been used, and the discharging time is typically less than a minute [16-19]. However, when connected to an external resistor, electrolytes with sluggish mass transfer can take nearly two hours for the current to stabilize from 6.2 mA to 3.9 mA, as reported by Wang et al. [16] in eutectic organic solvents. The inconsistent discharging conditions can result in significant inaccuracies in output power and thereby efficiency due to transient effects. At the steady state, the output electric current must be balanced with the internal ionic current, hence the output power density is determined by the rate-limiting process in the cell, either the reaction rate or more often the mass transfer rate. In transient measurements, however, the concentration of reactants at the electrode surface remains close to the initial state at very high scanning rates or small discharging duration, hence the measured current can be significantly higher than the steady-state limit. As we shall see later in this paper, power density, and efficiency can be significantly overestimated if the scanning rate is not chosen properly.

This work aims to establish an experimental protocol to properly quantify the energy conversion efficiency of LTC devices. We proposed a proper figure of merit $ZT = S_{tg}^2 T/R_{in} K$ for LTC devices, where $R_{in}$ and $K$ are the total internal resistance and thermal conductance across the LTC, respectively. The modified $ZT$ is derived by linearizing the Butler-Volmer relation describing the mass transfer overpotential and activation overpotential losses due to Faradaic charge transfer. The total internal resistance $R_{in}$ can be understood as a serial circuit containing mass transfer resistance, ohmic resistance, and charge transfer resistance. The modified $ZT$ based on device properties therefore naturally incorporated the electrochemical performance at electrode-electrolyte interfaces. We have also compared different methods for characterizing the output power, including constant resistance discharging, constant current steps discharging, and linear sweeping voltammetry (LSV). We found that LSV can significantly overestimate the output power and efficiency due to the transient effects, while constant resistance discharging and constant current steps discharging give consistent output power at the steady state. Heat conduction across the LTC is also carefully measured, and the natural convection of electrolytes makes a considerable parasitic contribution to heat conduction across the device. With careful experimental characterization, we showed that the modified $ZT = S_{tg}^2 T/R_{in}K$ is a proper efficiency indicator, provided that the steady state is ensured during measurements. Our work serves as a benchmark for quantifying efficiency for harvesting low-grade heat energy using LTCs.

## II. DERIVATION OF MODIFED ZT

We consider a generic LTC as shown in Fig. 1a operating under a temperature difference $\Delta T$ across the device. The redox reaction is written as $O + ne \leftrightarrows R$, where $O$ is the redox ion at the oxidized state and $R$ is the reduced state. $n$ is the number of electron transfers in the reaction. The effective thermopower of LTC is determined as $S_{tg} = -\Delta s_{rxn}/(nF)$ [1,14], with $F$ the Faraday constant. $\Delta s_{rxn} = s_R - s_O$ is the entropy difference between the species $R$ and $O$. The reactions directions are symmetric on the hot and the cold electrolyte. Take a p-type redox electrolyte as an example (as shown in Fig 1a). The entropy change for a p-type redox pair is negative, *i.e.* $\Delta s_{rxn} = s_R - s_O < 0$, hence the oxidized state $O$ has a higher entropy. Therefore, the anodic reaction consuming $R$ and generating $O$ is favored at the hot electrode, while the cathodic reaction converting $O$ to $R$ is thermodynamically favored at the cold side. For n-type redox electrolyte, cathodic and anodic at the two electrodes are reversed. The mass transfer across the LTC device ensures the continuous circulation of the reactants, such that the LTC can output current continuously under a constant temperature difference. At the steady state, the ionic current must be balanced with the Faradaic current at the electrochemical interface, therefore the output power is synergistically affected by mass transfer and reaction rate. This section presents a thermo-electrochemical model for computing the power density and efficiency of LTCs. In Section II.A, we outline the general equations determining the power density and efficiency. In Section II.B, the Butler-Volmer relation is linearized to determine the resistance originating from finite rates of mass transfer and reaction, by modeling a single electrochemical interface of a half-cell as shown in Fig. 1b. We proved that the effective internal resistance of the device can be understood as a serial connection of mass transfer resistances, charge transfer resistances, and the Ohmic resistance. In Section II.C, we show that the heat-to-electric conversion efficiency at steady-state is determined by a modified figure of merit $ZT = S_{tg}^2 T / R_{in} K$, where $R_{in}$ is the total internal resistance obtained by linearizing the Butler-Volmer relation, and $K$ is the thermal conductance of the device.

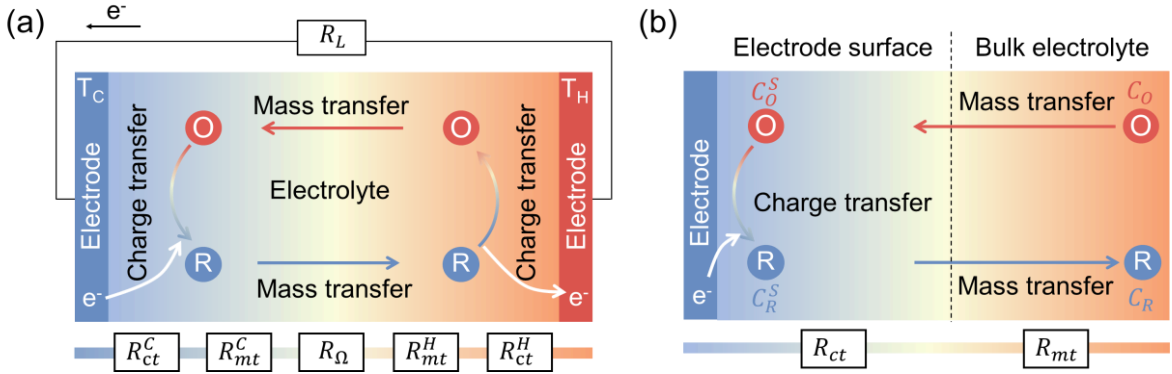

Fig. 1. (a) Continuous power output of an LTC device; (b) Mass transfer and electrochemistry in a half-cell reaction of LTC.

### A. Power density and efficiency of LTC

Consider an LTC connected to an external resistive load $R_L$, operating under a constant temperature difference $\Delta T$. The open-circuit voltage is determined as $V_{oc} = |S_{tg}|\Delta T$. When the system has a finite current output $I$, the voltage drops are due to a resistive voltage drop $IR_\Omega$ ($R_\Omega$ denotes the total ohmic resistance inside the LTC) and an overpotential voltage drop $\eta_{tot}$ due to electrode polarization. Therefore, the voltage-current relation is written as:

$$V(I) = |S_{tg}|\Delta T - (IR_\Omega + \eta_{tot}(I)) \qquad (1)$$

The voltage can also be expressed as $V(I) = IR_L$, and we can therefore express the current as:

$$I = \frac{|S_{tg}|\Delta T - \eta_{tot}(I)}{R_L + R_\Omega} \tag{2}$$

The output power of LTC is simply:

$$P = VI \tag{3}$$

To determine the efficiency, the heat absorbed from the reservoir is calculated as:

$$Q_H = |S_{tg}|T_H I + K\Delta T - \frac{1}{2}(I^2 R_\Omega + I\eta_{tot}) \tag{4}$$

Here, $|S_{tg}|T_H I$ is the reversible heat absorbed from the hot reservoir at a temperature $T_H$, $K\Delta T$ is the heat transfer across the LTC with $K$ the thermal conductance. The third term in Eq. (4) is the irreversible heat dissipated back into the hot reservoir, with $I^2 R_\Omega$ representing the Joule heat and $I\eta_{tot}$ denoting the irreversible polarization heat. Finally, the energy conversion efficiency is defined as:

$$\eta_E = \frac{VI}{Q_H} \tag{5}$$

The major difference between the LTC and solid-state TEG is the polarization overpotential $\eta_{tot}(I)$, which scales nonlinearly with the discharging current $I$. In addition, $\eta_{tot}(I)$ is a combined result of mass transfer and activation polarization (finite reaction rate). In the following Section II.B, we linearize the overpotential $\eta_{tot}(I)$ and show that the overpotential at a single electrode-electrolyte interface can be understood as the voltage drop across a serial connection of mass transfer resistance and charge transfer resistance.

**B. Effective resistance in linear polarization region**

Since the reactions are symmetric at the two electrodes, the overpotential losses can be effectively modeled simply by analyzing a half-cell reaction (Fig. 1b). The extended Butler-Volmer relation is used to describe a half-cell overpotential $\eta$ at finite current $I$:

$$I = I_0 \left[ \frac{C_O^S}{C_O} \exp\left(-\frac{\alpha F \eta}{\mathcal{R}T}\right) - \frac{C_R^S}{C_R} \exp\left((1-\alpha)\frac{F\eta}{\mathcal{R}T}\right) \right] \quad (6)$$

where $T$ is the electrode temperature, $C_O, C_R$ denotes the concentration of reactants $O$ and $R$ in the bulk electrolyte; $C_O^S$ and $C_R^S$ are the concentrations at the electrode surface, $\mathcal{R} = 8.314$ J/(mol·K) is the ideal gas constant, and $\alpha$ is the symmetry factor of the activation barrier for Faradaic charge transfer, typically ~0.5 for a reversible reaction [20]. $F = \mathcal{N}_A e$ is the Faraday's constant, with $\mathcal{N}_A$ the Avogadro's constant and $e$ the elementary charge. The $I_0$ is the exchange current at electrochemical equilibrium, determined as

$$I_0 = nFAk_0 C_O^{1-\alpha} C_R^{\alpha} \quad (7)$$

where $k_0$ the reaction rate constant [21] and $A$ the electrode area. In the Butler-Volmer relation, cathodic current is defined as positive ($I > 0$) where electrons are injected into the electrolyte, resulting in $O$ reduced to $R$. For a p-type redox pair as shown in Fig. 1, the cathode is the cold electrode kept at $T_C$ while the anode is the hot electrode. On the other hand, for an n-type redox pair, the cathode is the hot electrode and the anode is the cold electrode. Regardless of the sign of $S_{tg}$, the total overpotential loss is always determined as $\eta_{tot} = \eta_+ - \eta_-$, with $\eta_+$ ($\eta_-$) the overpotential at the cathodic (anodic) electrode. Such a definition ensures that $\eta_+ > 0$ and $\eta_- < 0$, and the total overpotential is always positive for LTCs.

At the steady state, the Faradic current is balanced with the ionic current to ensure the conservation of charges. Therefore, we can also express the current as:

$$I = nFAM_O(C_O - C_O^S) = -nFAM_R(C_R - C_R^S) \quad (8)$$

where $M_O$ and $M_R$ are mass transfer coefficient of reactants $O$ and $R$ respectively. We can therefore eliminate the surface concentrations $C_O^S$ and $C_R^S$, and the Butler-Volmer relation can be rewritten as:

$$\frac{I}{I_0} = \left(1 - \frac{I}{nFM_O C_O}\right)\exp\left(-\alpha\frac{F\eta}{\mathcal{R}T}\right) - \left(1 + \frac{I}{nFM_R C_R}\right)\exp\left((1-\alpha)\frac{F\eta}{\mathcal{R}T}\right) \quad (9)$$

For LTCs, the operating temperature difference $\Delta T$ is typically a few dozen Kelvin, such that the open-circuit voltage is on the order of 0.1 V [22-24]. Given an external load of ~ 1 Ω, the current density is also only a few mA/cm², such that the polarization typically scales linearly with $I$ [24,25]. We can therefore linearize the exponential factors $\exp\left(-\alpha\frac{F\eta}{\mathcal{R}T}\right) \approx 1 - \alpha\frac{F\eta}{\mathcal{R}T}$ and $\exp\left((1-\alpha)\frac{F\eta}{\mathcal{R}T}\right) \approx 1 + (1-\alpha)\frac{F\eta}{\mathcal{R}T}$. After some algebra, Eq. (9) can be simplified as:

$$\eta = I\frac{\mathcal{R}T}{F}\left(\frac{1}{I_0} + \frac{1}{nFAM_O C_O} + \frac{1}{nFAM_R C_R}\right) \quad (10)$$

The half-cell polarization resistance can therefore be defined as:

$$R_{pol}^{1/2} = \frac{\eta}{I} = \frac{\mathcal{R}T}{F}\left(\frac{1}{I_0} + \frac{1}{nFAM_O C_O} + \frac{1}{nFAM_R C_R}\right) \quad (11)$$

We can separate the half-cell polarization resistance into charge transfer resistance $R_{ct}$ and mass transfer resistance $R_{mt}$:

$$R_{pol}^{1/2} = R_{ct} + R_{mt} \quad (12)$$

$$R_{ct} = \mathcal{R}T/(I_0 F) \quad (13)$$

$$R_{mt} = \frac{\mathcal{R}T}{F}\left(\frac{1}{nFAM_O C_O} + \frac{1}{nFAM_R C_R}\right) \quad (14)$$

Therefore, the half-cell polarization resistance can be understood as a serial connection of $R_{ct}$ and $R_{mt}$ in the linear polarization region, as shown in Fig. 1b.

For the entire LTC, the total overpotential loss happens both at the hot and the cold electrodes, hence the total polarization resistance is therefore:

$$R_{pol} = R_{pol}^{1/2}(T_H) + R_{pol}^{1/2}(T_C) = R_{ct}^H + R_{mt}^H + R_{ct}^C + R_{mt}^C \quad (15)$$

where $R_{ct}^H$ and $R_{ct}^C$ are the half-cell charge transfer resistance evaluated at $T_H$ and $T_C$, and $R_{mt}^H$ and $R_{mt}^C$ correspond to the mass transfer resistances at the hot and cold electrodes, respectively. Substituting $\eta_{tot} = IR_{pol}$ into Eq. (1), we can obtain:

$$V(I) = |S_{tg}|\Delta T - I \cdot R_{in} \tag{16}$$

where $R_{in} = R_{pol} + R_\Omega$ is the effective internal resistance. We have therefore derived the effective circuit of internal resistance for an LTC device, as shown in Fig. 1a.

### C. Maximum efficiency and the figure of merit

After linearizing the voltage-current relation, we can substitute Eq. (16) into Eq. (5), and the efficiency can be expressed as a function of $r = R_L/R_{in}$:

$$\eta_E(r) = \frac{(1 - T_C/T_H)r}{(1+r) - \frac{1}{2}(1 - T_C/T_H) + (1+r)^2 \frac{T_C}{T_H}/Z\bar{T}} \tag{17}$$

where $\bar{T} = (T_H + T_C)/2$ is the average temperature and $Z\bar{T}$ is the figure of merit for LTCs:

$$Z\bar{T} = \frac{S_{tg}^2 \bar{T}}{R_{in} K} \tag{18}$$

The maximum efficiency is attained when $d\eta_E/dr = 0$, and the optimal load is determined as $r = R_L/R_{in} = \sqrt{Z\bar{T} + 1}$. The maximum efficiency therefore takes the same form as the solid-state thermoelectrics:

$$\eta_{E,max} = \left(1 - \frac{T_C}{T_H}\right) \frac{\sqrt{1 + Z\bar{T}} - 1}{\sqrt{1 + Z\bar{T}} + \frac{T_C}{T_H}} \tag{19}$$

It is necessary to note that the figure of merit $Z\bar{T}$ of LTC is only well-defined using the internal resistance and thermal conductance of the device. The internal resistance is a combined result of activation polarization, mass transfer, and Ohmic resistance of the LTC, therefore defining the figure of merit using electrolyte properties is meaningless. As we shall see in the derivation, the figure of merit $Z\bar{T}$ is only applicable at the steady state and within the linear polarization region. If the discharging voltage-current curve $V(I)$ shows significant nonlinearity, the figure of merit is not applicable. The $Z\bar{T}$ should also not be used in transient conditions. For some ionic thermoelectric capacitors with redox-inert ions, the discharging is transient and capacitive, the $Z\bar{T}$ derived here should not be used for efficiency evaluation either.

In addition to Eq. (19), the maximum efficiency of LTC can also be determined by directly measuring the harvested power at the optimal load. Due to the low ionic mobility, the ohmic internal resistance $R_\Omega$ is typically a few Ω, the current is only a few mA. In this case the reversible heat $|S_{tg}I|T_H$ is therefore typically a few mW ($T_H < 373$ K), and the irreversible heat generation $I^2 R_{in}$ is also negligibly small (a few μW), compared with the heat conduction $K\Delta T$ (0.1 W). Therefore, the heat absorbed from the hot reservoir is dominated by heat transfer across the device $Q_H \approx K\Delta T$. This allows us to easily the maximum efficiency approximately at maximum power [22,26,27]:

$$\eta_{E,max} = \max[P/Q_H] \approx \frac{P_{max}}{K\Delta T} \tag{20}$$

This approximation is valid because the operating current is small, such that heat conduction dominates over heat generation in LTC. For solid-state thermoelectrics with large operating currents (*e.g.* a few A), the maximum efficiency and the efficiency at the maximum power are not equal. To prove the validity of the modified $Z\bar{T}$, we will experimentally measure $S_{tg}$, $R_{in}$, $K$, and $P_{max}$, and compare the optimal efficiencies evaluated through $Z\bar{T}$ by Eq. (19) and $P_{max}$ through Eq. (20).

### III. EXPERIMENTAL METHODS

This section details the experimental methods for determining the maximum power and efficiency of LTCs. Section 3.1 discusses the preparation of materials and fabrication of the LTC device. Section 3.2 describes methods for measuring thermal voltage and discharging voltage-current curves under a finite temperature difference $\Delta T$. Section 3.3 focuses on measuring the thermal conductance of the LTC.

### A. Materials preparation and device fabrication

We use the well-studied redox pair $Fe(CN)_6^{3-}/Fe(CN)_6^{4-}$ as the thermogalvanic electrolyte. The aqueous electrolyte (0.4 M) is prepared by dissolving potassium ferricyanide ($K_3Fe(CN)_6$, 99.5%) and potassium ferrocyanide ($K_4Fe(CN)_6$, 99%) in deionized water with a 1:1 molar ratio. We further added 3M guanidinium chloride (GdmCl, 99%) into aqueous $Fe(CN)_6^{3-}/Fe(CN)_6^{4-}$ electrolyte to induce thermosensitive crystallization between $Gdm^+$ and $Fe(CN)_6^{4-}$. Such thermosensitive crystallization can boost the concentration gradient across the device, resulting in an improved thermopower [9]. Carbon paper is used as electrode and ultrasonically cleaned in acetone, ethanol, and distilled water for five minutes each to remove the surface contaminants. To enhance the hydrophilicity of the surface, carbon papers are heated to 500°C for 12 hours under ambient atmosphere in a tube furnace. Then we prepare a catalyst slurry by mixing platinum-carbon slurry with polyvinylidene fluoride (PVDF) binder in a 9:1 mass ratio, using N-methyl-2-pyrrolidone (NMP, 99.9%) as the solvent. The slurry was coated onto the surface of the carbon paper and then dried in a vacuum oven at 60°C for 12 hours.

After preparing the electrodes and electrolytes, the LTC device is assembled as shown in Fig. 2. A polymeric insulation shell, two carbon paper electrodes, and two graphite current collectors are stacked together. To prevent leakage, silicone gaskets are placed between electrodes and current collectors, and all components are pressurized by two pieces of stainless steel caps using nuts and bolts. Such assembly creates a cavity to contain the thermogalvanic electrolyte. The insulation shell is made from polytetrafluoroethylene (PTFE) with a pinhole for injecting electrolytes into the device. We employ two Peltier modules in contact with the stainless steel caps to create a temperature difference $\Delta T$ across the device. Programmable power sources are used to supply currents to the Peltier modules to heat or cool the device, and the temperatures are regulated using a PID controller. Thermocouples are attached to the current collectors to measure the temperature difference across the device.

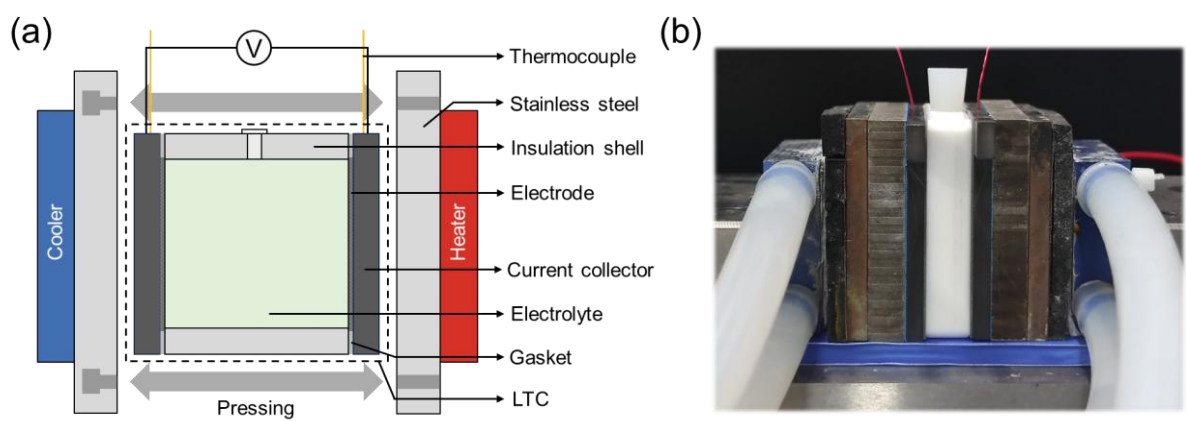

Fig. 2. (a) Schematic and (b) photo of the LTC device.

**B. Thermopower and discharging measurements**

An electrochemical workstation (CS310X) was used to characterize the open-circuit thermal voltages, discharge currents, and voltages. The thermopower $S_{tg}$ is measured by increasing $\Delta T$ across the device in a few steps, and the open-circuit voltage is measured after the temperature and voltage are stabilized at each step, as shown in Fig. 3a. The thermopower can be easily obtained from the slope of voltage-$\Delta T$ relation through linear regression. As shown in Fig. 3b, the measured $S_{tg}$ of aqueous $Fe(CN)_6^{3-}/Fe(CN)_6^{4-}$ is 1.39 mV/K, and the thermopower of $Fe(CN)_6^{3-}/Fe(CN)_6^{4-}$+$Gdm^+$ with thermosensitive crystallization is enhanced to 3.52 mV/K, showing nice agreement with previous reports [9]. To determine the ohmic resistance of the cell, we make electrochemical impedance spectroscopy (EIS) from 100 kHz to 0.1 Hz with an amplitude of 10 mV. The Ohmic resistance is determined as 4.04 Ω when the EIS curve intersects with the real axis in the Nyquist plot. Charge transfer resistance $R_{ct}$ is determined as 0.1 Ω from the diameter semicircular trajectories of the impedance.

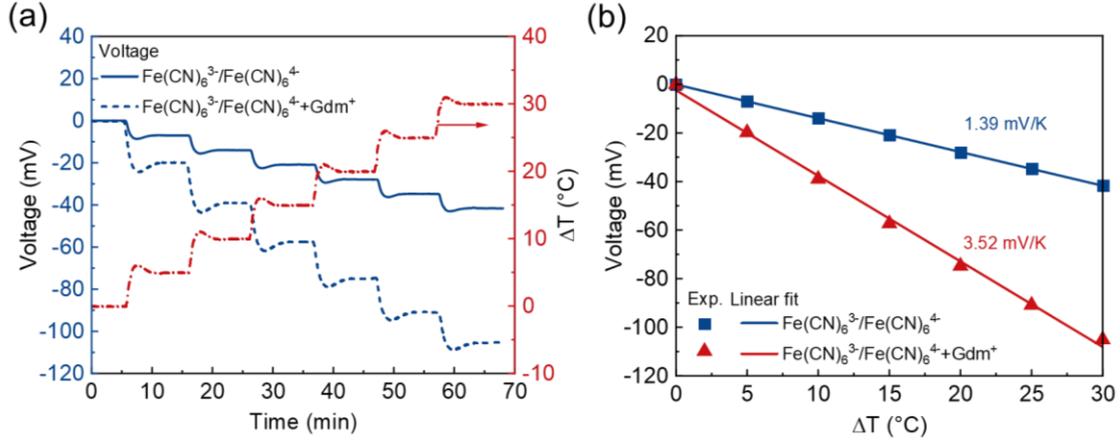

Fig. 3. (a) Thermal voltage of LTCs using $Fe(CN)_6^{3-}/Fe(CN)_6^{4-}$ and $Fe(CN)_6^{3-}/Fe(CN)_6^{4-}+Gdm^+$ in response to the applied temperature difference; (b) Effective thermopower $S_{tg}$ extracted from the temperature dependent thermal voltages. Voltage is measured as the electrode potential difference $E(T_H) - E(T_C)$.

We compared and tested several commonly used methods for measuring the discharge behavior, including constant current steps, constant resistance discharging, and linear sweep voltammetry (LSV). In the constant current step method, a series of increasing discharging currents are applied in a stepwise manner, and the responsive voltage $V(I)$ is measured at each current step, as shown in Fig. 4a. Each current step lasted ~120 s to ensure the voltage is fully stabilized. Constant resistance discharging is very straightforward. Both electrodes of LTC are connected to an external resistor $R_L$, then the voltage and current can be recorded, as shown in Fig. 4b. Fig. 4c shows the commonly used LSV method to determine the discharging curve $V(I)$. LSV applies a linearly varying voltage at a controlled scanning rate and measures the resulting current. Note that LSV is an intrinsically transient measurement, and as we shall see in Section IV, the scanning rate of LSV must be carefully chosen to avoid overestimated discharging power.

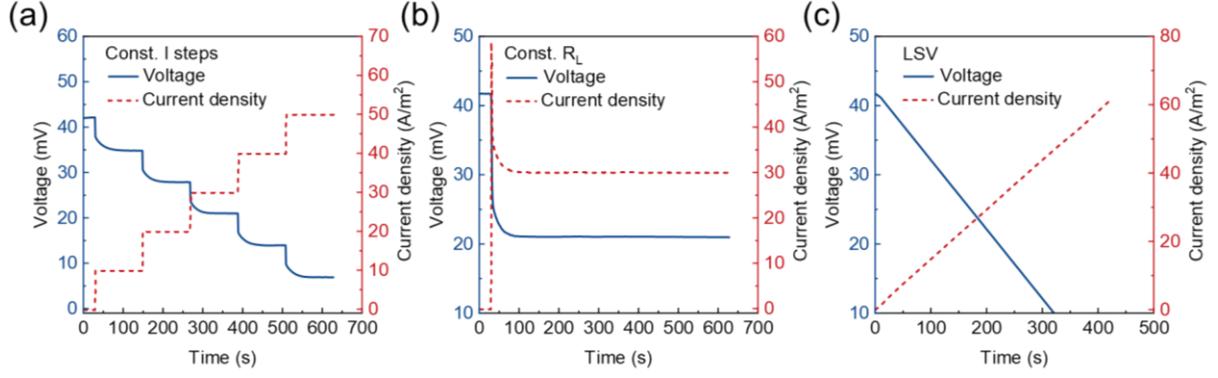

Fig. 4. Discharging voltage and current using (a) constant current steps, (b) constant resistance discharging, and (c) LSV method.

**C. Thermal conductance measurement**

We use the steady-state method to determine the thermal conductance $K$ of LTC [9,17], as shown in Fig. 5a. The LTC device is in contact with a reference PTFE plate, and surrounded by thermal-insulating polystyrene foam to ensure one-dimensional heat transfer. When a temperature difference $\Delta T$ is applied across the testing state, we use an infrared camera (FLIR A325sc) to measure the temperature profile across the reference PTFE plate and the LTC device, as shown in Fig. 5b. At the steady state, the heat transfer across the PTFE should be equal to that across the LTC device, hence the heat transfer rate is calculated by Fourier's law:

$$Q_{LTC} = k_{PTFE} A_C \left(\frac{\Delta T}{L}\right)_{PTFE} \qquad (21)$$

where $k_{PTFE}$ is the thermal conductivity of the PTFE plate. Using the hot-wire method, $k_{PTFE}$ is measured as 0.256 W/(m·K), agreeing well with other literature reports [28,29]. $A_C$ and $L$ are the cross-section area (1 cm$^2$) and the thickness (5 mm) of the PTFE reference sample, respectively. The thermal conductance $K$ of the LTC device was then calculated as:

$$K = \frac{Q_{LTC}}{\Delta T_{LTC}} \qquad (22)$$

where $\Delta T_{LTC}$ is the temperature difference across the LTC's current collectors. As shown in Fig. 5b, the parallel temperature contour indicates that the heat transfer is largely one-dimensional along the horizontal direction. The thermal conductance $K$ of the LTC device is determined as 0.0084 W/K.

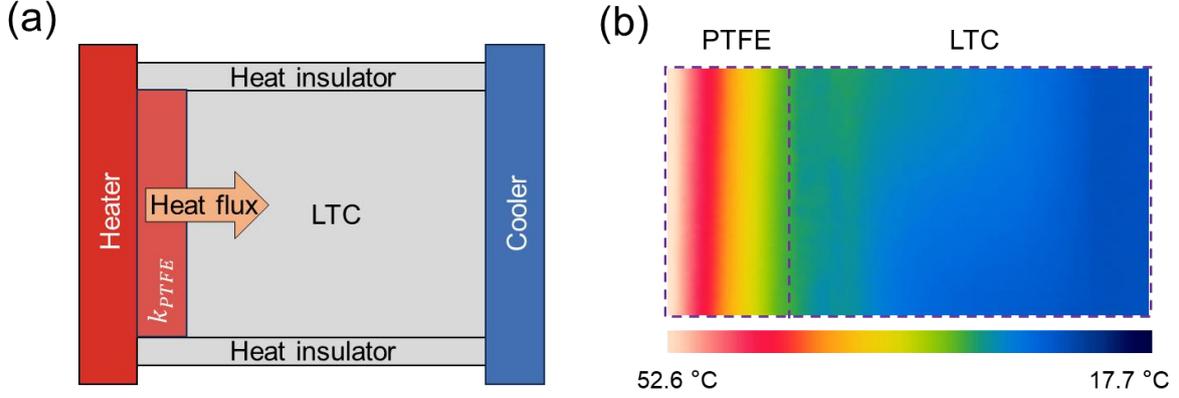

Fig. 5. (a) Diagram of the thermal conductance measurement setup for the LTC; (b) Temperature profile across the testing device using the infrared camera.

## IV. MEASUREMENT RESULTS

Thermo-electrochemical measurements results are organized as follows. In Section IV.A, we demonstrate the validity of using the modified $Z\bar{T}$ to quantify efficiency. We showed that the following two methods give consistent results: (i) first measure the modified device $Z\bar{T}$ at steady state, and then calculate the efficiency using Eq. (19); and (ii) measure $P_{max}$ at the steady state and take the ratio to the heat transfer rate $K\Delta T$. Section IV.B compares different discharging methods such as constant current steps, constant resistance discharging, and linear sweeping voltammetry (LSV). We show that the transient effects will lead to severely overestimated power density. Section IV.C highlights the importance of accurately measuring the thermal conductance $K$. Natural convection significantly affects the device conductance $K$, and we show that estimating $K$ either using the thermal conductivity of the electrolyte or through the total heat conduction resistance would result in overestimated efficiencies.

## A. Validity of the modified ZT

We take a two-step process to show the validity of the modified $Z\bar{T} = S_{tg}^2 \bar{T}/(R_{in}K)$. First, we testify to the linear polarization assumption, which is the prerequisite for the internal resistance $R_{in}$ to be well-defined. Then, we compare the efficiency measured through output power and heat transfer rate $P_{max}/(K\Delta T)$ (referred to as the measured efficiency) with the efficiency evaluated using the figure of merit $Z\bar{T}$.

As shown in Fig. 6a, we have measured the voltage-current relation using the constant current steps method at a temperature difference $\Delta T = 30$ K. We made sure that each current step is long enough for the responsive voltage to fully stabilize. The voltage-current discharging curve showed excellent linearity, allowing us to determine the internal resistance as $R_{in} = V_{oc}/I_{sc}$, with $V_{oc}$ the open-circuit voltage and $I_{sc}$ the short-circuit current. The internal resistance of the LTC device is estimated as 7 Ω. The measured voltage-current curve agrees well with the linearized discharge model $V(I) = V_{oc} - IR_{in}$. The discharging $V(I)$ is also calculated by fully solving the Butler-Volmer relation as the benchmark. The consistency of measurements with the linearized discharge model validated the linear polarization approximation. The maximum power density can be easily identified as 0.6 W/m². However, when the transient LSV is used, the short circuit current is 68.3% larger than the results from constant current steps. While the discharging curve is still linear, the internal resistance $R_{in}^t$ at the transient state is only 4.1 Ω.

Fig. 6b compares the efficiency obtained from discharging tests in Fig. 6a. Using the constant current steps, the measured efficiency is determined as 0.27% relative to the Carnot limit. With $R_{in} = 7$ Ω and $K = 0.0084$ K/W, the steady-state $Z\bar{T}$ is determined as 0.01, and the Carnot relative efficiency is evaluated as 0.27 % using Eq. (19), showing excellent agreement with the measured value. However, when transient LSV discharging (100 mV/s) is used, the peak power density is overestimated to 1.1 W/m², and the measured efficiency at the transient state is nearly doubled compared with the steady-state result. Similarly, the transient figure of merit $Z_t\bar{T} = S_{tg}\bar{T}/(R_{in}^t K)$ is 77.1% higher than the steady-state value, and the efficiency using $Z_t\bar{T}$ is also overestimated. Finally, efficiencies are calculated using $Z_{el}\bar{T}$ purely based on electrolyte properties shows the most severe overestimation, due to the neglected mass transfer resistance and charge transfer resistance near electrode surfaces. This comparison justified the validity of using the modified $Z\bar{T}$ as the efficiency indicator, as long as the steady state discharging is ensured. We also demonstrate that the efficiency can be severely overestimated when the transient discharging test is used or when $Z\bar{T}$ is calculated inappropriately only using electrolyte properties.

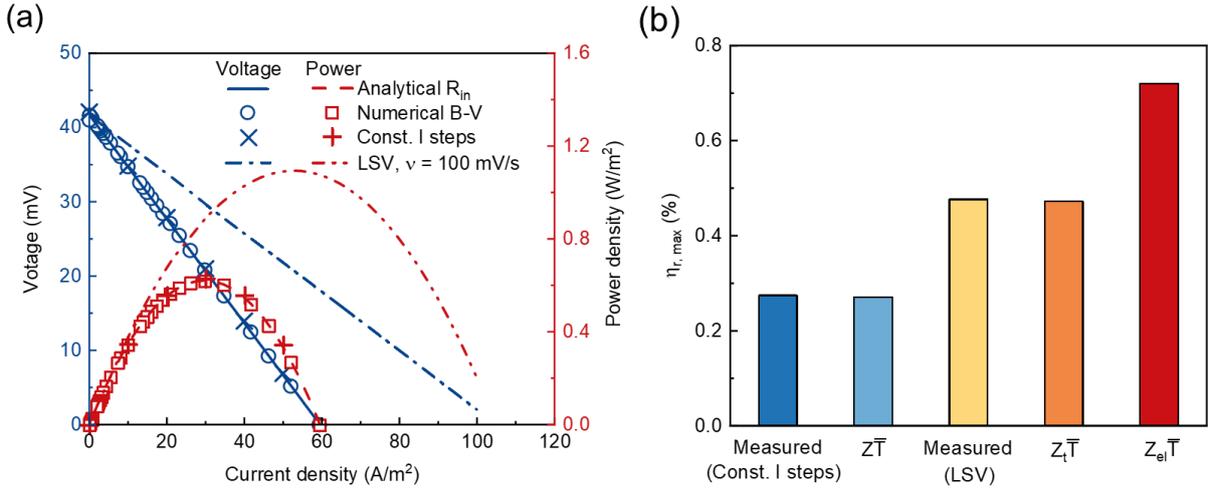

Fig. 6. (a) Comparison between the modeled and measured discharging voltage (power)-current curves of LTC with $\Delta T = 30$ K. The modeled discharging curves are calculated using the analytical linearized model using the effective internal resistance, and by numerically solving the Butler-Volmer (B-V) relation. Measurements are performed using constant current steps (const. $I$ steps) and LSV (scanning rate 100 mV/s). (b) Comparison of the maximum efficiency of LTC determined by different methods. The "measured" efficiency is determined by taking the ratio between the maximum power and the heat transfer rate, with the power measured by either LSV or constant current steps (const. $I$ steps). Efficiencies labeled by $Z\bar{T}$, $Z_t\bar{T}$, and $Z_{el}\bar{T}$ are calculated using Eq. (19). $Z_t\bar{T} = S_{tg}^2 \bar{T}/R_{in}^t K$ is determined based on $R_{in}^t$ obtained from transient LSV. $Z_{el}\bar{T} = S_{tg}^2 \sigma_{el}\bar{T}/\kappa_{el}$ is determined using electrolyte conductivity and thermal conductivity. All efficiencies are normalized by the Carnot limit $1 - T_C/T_H$.

## B. Transient effects on power and efficiency

This section investigates the transient effects on the measured output power and efficiency of LTC devices. Fig. 7 summarizes the measurement results of $Z\bar{T}$, power density $P_{max}/A$, and the Carnot-relative efficiency $\eta_{r,max} = \eta_{E,max}/(1 - T_C/T_H)$ using different discharging methods, including constant resistance discharging, constant current steps, and LSV. The constant resistance discharging time, the duration of each constant current step, and the scanning rates for LSV are selected based on reported literature [16-19,24,27,30]. The $Z\bar{T}$, power density, and efficiency measured using a 2-minute constant resistance discharging were consistent with those acquired over 10 minutes, indicating that a 2-minute duration is sufficiently long to ensure voltage stabilization. Similarly, when constant current steps are used, one has to ensure that each current step must be maintained long enough for the voltage to stabilize. For example, when only using a 5-second duration for each current step, the measured power density is 30 % higher than that obtained using longer durations. When the duration of each current step is longer than 2 minutes, the testing results become consistent with constant resistance discharging. We indicate this steady-state discharging using a dashed horizontal line in Fig. 7. For LSV discharging, only the lowest scanning rate of 0.1 mV/s can ensure steady-state behavior, while high scanning rates of 4.2 ~ 32.5 mV/s significantly overestimated the device performance.

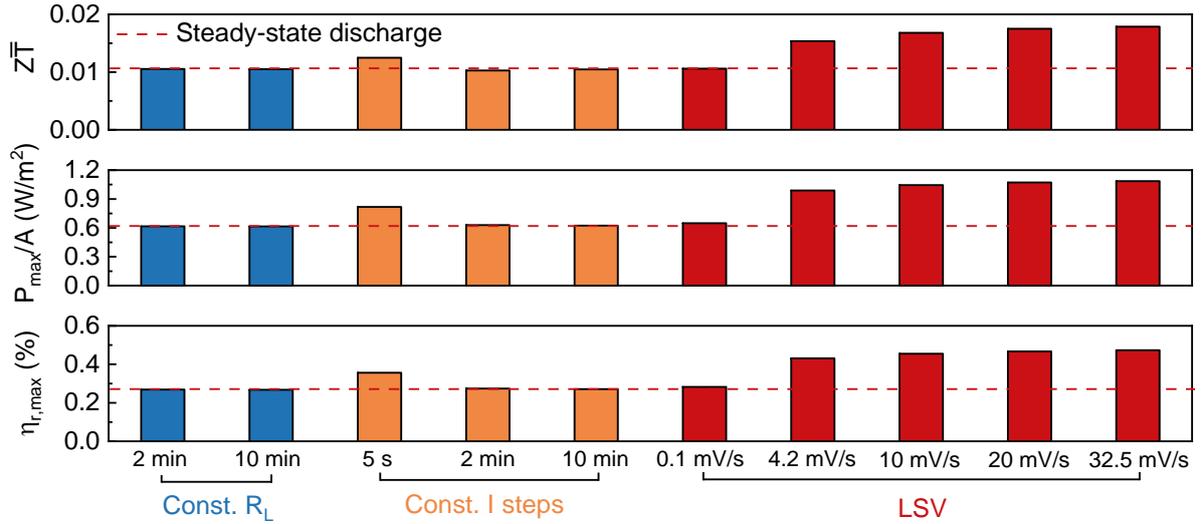

Fig. 7. The measured figure of merit, maximum power density, and Carnot relative efficiency using constant resistance discharging (const. $R_L$), constant current steps (const. $I$ steps), and linear sweeping voltammetry (LSV).

Fig. 8 further illustrates the scanning-rate dependence in LTC performance in LSV measurements. As the scanning rate $v$ increases from 0.025 mV/s to 100 mV/s, a nearly one-fold increase in normalized power density is observed from 0.7 mW/(m$^2$·K$^2$) to 1.2 mW/(m$^2$·K$^2$) for pristine Fe(CN)$_6^{3-}$/Fe(CN)$_6^{4-}$ (Fig. 8a). When the scanning rate is sufficiently fast, there is no time for the concentration boundary layer to be fully developed, such that the mass transfer resistance is suppressed. For the Fe(CN)$_6^{3-}$/Fe(CN)$_6^{4-}$+Gdm$^+$ with thermosensitive crystallization effects, the scanning rate dependence is much more pronounced, with the normalized power density $P_{max}/(A\Delta T^2)$ increased by an order of magnitude from 0.4 mW/(m$^2$·K$^2$) to 6 mW/(m$^2$·K$^2$), with the scanning rate increased from 0.025 mV/s to 100 mV/s. When the scanning rate is below 0.25 mV/s, $P_{max}/(A\Delta T^2)$ of Fe(CN)$_6^{3-}$/Fe(CN)$_6^{4-}$+Gdm$^+$ is even lower than the pristine Fe(CN)$_6^{3-}$/Fe(CN)$_6^{4-}$ despite its high thermopower, possibly due to the large mass transfer resistance. We also included measurement results from literature reporting the scanning rate of LSV. For Fe(CN)$_6^{3-}$/Fe(CN)$_6^{4-}$+Gdm$^+$, the measured power by Wang et al. [16] showed a nice agreement with our results. For the pristine aqueous $Fe(CN)_6^{3-}$/$Fe(CN)_6^{4-}$, $P_{max}/(A\Delta T^2)$ reported by Duan et al. [17] and Yu et al. [18] is much lower than ours, possibly due to the lower catalytic activity of the graphite electrode the authors used. Nevertheless, the increased power density measured at higher scanning rates is consistent with the trend observed in this work. Normalized power density $P_{max}/(A\Delta T^2)$ of pristine $Fe(CN)_6^{3-}$/$Fe(CN)_6^{4-}$ converged to steady-state limit when the scanning rate is lower than 0.1 mV/s, and the measured voltage-current relation agrees well with constant resistance discharging and constant current steps, as shown in Fig. 8b. This indicates that when using LSV for discharging tests, it is necessary to test the scanning rate dependence to make sure that the steady-state is reached. In Fig. 8c, the measured efficiency and the figure of merit of $Fe(CN)_6^{3-}$/$Fe(CN)_6^{4-}$ also show a pronounced dependence on the scanning rate of LSV. When the scanning rate is sufficiently low, both efficiency and the $Z\bar{T}$ can be reasonably characterized. At the fast scanning rate limit, however, the mass transfer resistance is severely suppressed, resulting in severely overestimated performance. For $Fe(CN)_6^{3-}$/$Fe(CN)_6^{4-}$+$Gdm^+$ with thermosensitive crystallization, we found that it is very hard to reach steady-state discharging. As shown in Fig 8d, the voltage, current, and output power keep decreasing even after discharging for an hour. In these cases, one should not use the $Z\bar{T}$ to characterize the efficiency, since the figure of merit is only well-defined for steady-state power output.

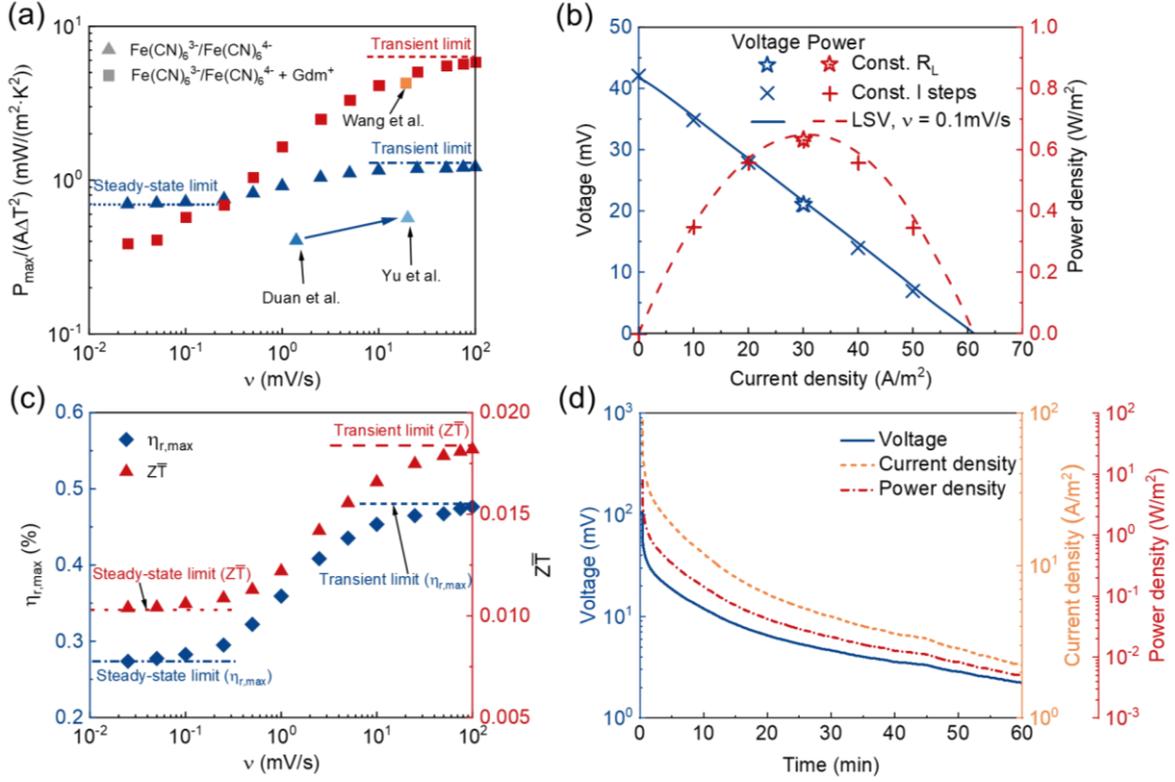

Fig. 8. (a) Normalized maximum power density $P_{max}/(A\Delta T^2)$ of the pristine aqueous $Fe(CN)_6^{3-}/Fe(CN)_6^{4-}$ and the $Fe(CN)_6^{3-}/Fe(CN)_6^{4-}+Gdm^+$ electrolyte with thermosensitive crystallization at different LSV scanning rates. Reference data for $Fe(CN)_6^{3-}/Fe(CN)_6^{4-}$ were obtained from Duan et al. [17] and Yu et al. [18], and reference data for $Fe(CN)_6^{3-}/Fe(CN)_6^{4-}+Gdm^+$ was reported by Wang et al. [16]; (b) Discharging voltage and power of aqueous $Fe(CN)_6^{3-}/Fe(CN)_6^{4-}$ measured by LSV (0.1 mV/K), compared with constant resistance discharging and constant current steps. (c) Scanning rate dependence of $Z\bar{T}$ and maximum Carnot-relative efficiency $\eta_{r,max}$ of aqueous $Fe(CN)_6^{3-}/Fe(CN)_6^{4-}$; (d) Temporal evolution of voltage, current, and power during a constant resistance discharge (10 Ω) of $Fe(CN)_6^{3-}/Fe(CN)_6^{4-}+Gdm^+$ electrolyte.

## C. Heat transfer across LTC

Heat transfer across LTC devices is analyzed in this part. The thermal resistance network of LTC is shown in Fig. 9, including resistances of the current collectors, electrodes, and the electrolyte. In the electrolyte, there will be natural convection driven by the temperature difference, affecting the electrolyte thermal resistance. In Fig. 9a, we have neglected the contact resistances between electrode and current collectors, and between electrode and electrolyte. Typical Kapitza length $L_K = \kappa R_c$ between at a liquid-solid interface is ~ 10 nm, and the pressurized solid-solid interface is in the micrometer range [31,32]. In comparison, the liquid electrolyte layer is 1 cm thick with a low thermal conductivity of 0.55 W/(m·K), hence the interface resistances are negligibly small for macroscale devices. The conduction resistance is estimated by first measuring the thermal conductivity of each component, and then using the resistance network in Fig. 9a to obtain the total thermal resistance. As shown in Fig. 9b, the total conduction resistance of the device is dominated by liquid electrolytes. However, the measured $K$ of LTC is nearly 50% higher than the conductance considering heat conduction only. As a result, the efficiency could also be severely overestimated if the natural convection of the electrolyte is neglected (Fig. 9c).

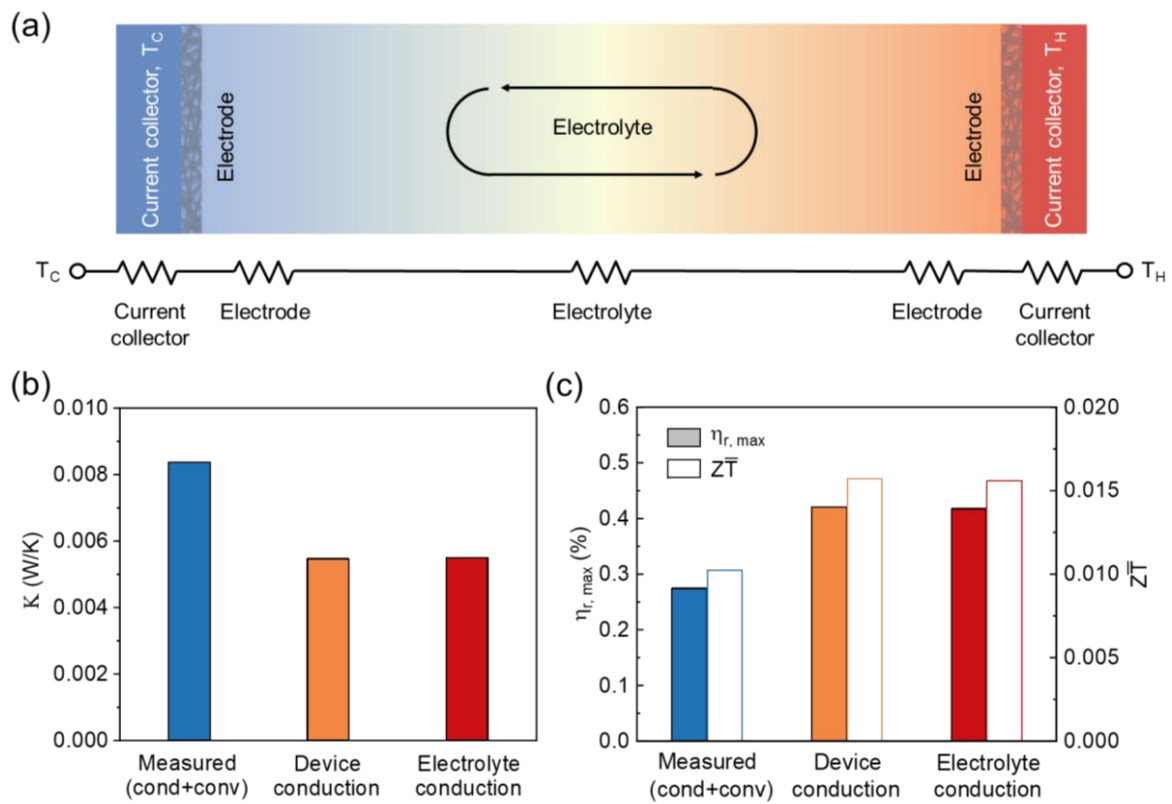

Fig. 9. (a) Thermal resistance network in LTC; (b) Measured thermal conductance of LTC, and the conductance of heat conduction across the entire device and the electrolyte only; (c) Figure of merit $Z\bar{T}$ and the maximum Carnot-relative efficiency $\eta_{r,max}$ evaluated using different thermal conductance values in (b).

Fig. 10 further shows the orientational dependence of natural convection. We have measured thermal conductance and discharging behavior for the following three cases: (i) $\Delta T$ is applied horizontally; (ii) $\Delta T$ is applied vertically, with the cold electrode above the hot electrode (cold-over-hot), and (iii) the hot-over-cold configuration where the hot electrode is above the cold one. For the hot-over-cold orientation, natural convection is severely suppressed, and the measured thermal conductance is the smallest (0.0051 W/K). This result agrees well with the value estimated using the conduction model (0.0055 W/K), suggesting that the natural convection has vanished. However, the absence of natural convection leads to severely suppressed output power, which is only 33.3% of the cold-over-hot configuration. As a result, the hot-over-cold configuration also shows the lowest efficiency among the three cases. For the other two configurations, despite the higher thermal conductance, natural convection helps to accelerate the mass transfer, and the measured power is much higher than the hot-over-cold configuration. Our observation agrees well with Kang et al. [33]. This characterization demonstrated the critical role of natural convection in affecting the power output and heat transfer across the device, and highlights the necessity of directly measuring the thermal conductance $K$.

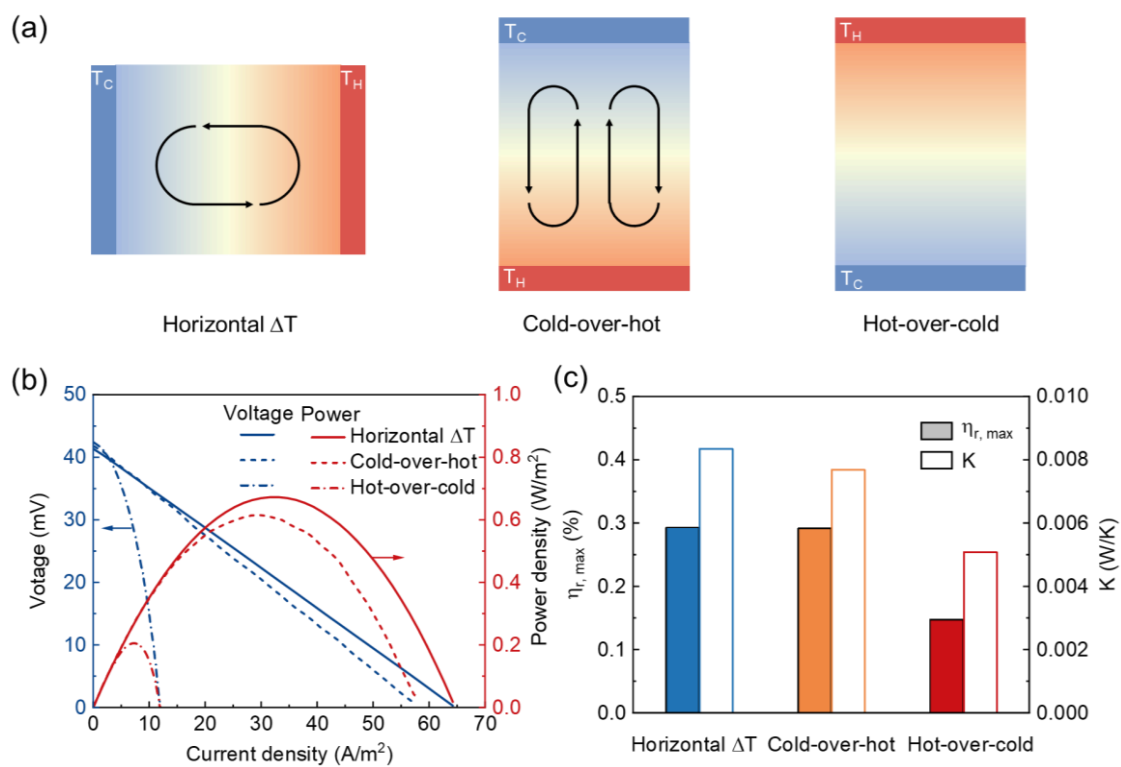

Fig. 10. (a) Different orientations of applied $\Delta T$; (b) Discharging voltage and current of $Fe(CN)_6^{3-}/Fe(CN)_6^{4-}$ with different orientations of $\Delta T = 30$ K, measured using LSV at 0.02 mV/K; (c) Thermal conductance $K$ and corresponding efficiency $\eta_{r,max}$ of LTCs with the three temperature gradient orientations.

## V. SUMMARY AND CONCLUSIONS

We have thoroughly investigated factors that can affect accuracy in characterizing the efficiency of LTCs for harvesting low-grade heat energy. Major contributions of this work, together with guidelines for properly quantifying the efficiencies of LTCs are summarized as follows:

(1) We demonstrated that the figure of merit $Z\bar{T} = S_{tg}^2 \bar{T}/(R_{in}K)$ of LTC is only well-defined based on device resistance $R_{in}$ and thermal conductance $K$, and can serve as a valid efficiency indicator for steady-state power output. The $Z\bar{T}$ defined in this work included the coupled effects of mass transfer and reaction kinetics near the electrode-electrolyte interfaces. Using electrolyte properties to quantify the figure of merit will result in a severe overestimation of efficiency.

(2) Efficiency characterization results are very sensitive to discharging methods. Steady-state power output can be safely ensured when using constant resistance discharging or constant current steps. In contrast, the commonly used LSV tends to overestimate output power even by an order of magnitude, if the scanning rate is not carefully selected. When using LSV, one has to measure the power output across a wide range of scanning rates, and make sure that the scanning rate is sufficiently low to prevent overestimated performances due to transient effects.

(3) It is crucial to experimentally measure the device's thermal conductance because the natural convection of electrolyte can contribute to nearly 1/3 of the device conductance $K$. Natural convection also plays a crucial role in supplying current to the electrochemical interfaces, as demonstrated by applying the temperature difference in different orientations.

**Acknowledgments**

Xin Qian and Ronggui Yang acknowledge support from the National Natural Science Foundation of China (NSFC Grant No. 52276065 and Grant No. 52036002). The authors declare no conflict of interest.